\documentclass[twocolumn,showpacs,preprintnumbers,amsmath,amssymb,prl,superscriptaddress]{revtex4}

\usepackage{graphicx}
\usepackage{dcolumn}
\usepackage{bm}

\begin{document}

\preprint{}

\title{Mechanism of Terahertz Electromagnetic Wave Emission from \\
Intrinsic Josephson Junctions}

\author{Masashi Tachiki}
\affiliation{
Graduate school of Frontier Science, The University of Tokyo, 5-1-6 Kashiwanoha, Kashiwa 277-8581, Japan}
\author{Shouta Fukuya}
\affiliation{%
Institute for Solid State Physics, University of Tokyo, Kashiwa, Chiba 277-8581, Japan}
\author{Tomio Koyama}
\affiliation{
Institute for Materials Research, Tohoku University, Katahira 2-1-1, Aoba-ku, Sendai 980-77, Japan}%

\date{\today}

\begin{abstract}
Using a 3-D parallelepiped model of intrinsic Josephson junctions, we calculate the cavity resonance modes of Josephson plasma waves excited by external electric currents.  The electromagnetic (EM) wave of the excited Josephson plasma is converted to a THz EM wave at the sample surfaces.  The cavity modes accompanied by static phase kinks of the superconducting order parameter have been intensively investigated.  The phase kinks induce a spatial modulation of the amplitude of the order parameter around the kinks and decrease the superconducting condensation energy.  The Josephson plasma produces a magnetic field in the vacuum in addition to the emitted EM wave.  This magnetic energy detemines the orientation of the cavity mode.  Taking account of the facts mentioned above, we obtained sharp resonance peaks in the I-V curves and sizable powers of continuous and coherent terahertz wave emission at the cavity resonance.  The emission frequencies are inversely proportional to the length of the shorter side of the samples in agreement with experiments.
\end{abstract}

\maketitle
Emission of terahertz electromagnetic (EM) waves from intrinsic Josephson junctions (IJJ) in high temperature superconductors has been extensively studied \cite{klei1992,sakai1993,koyama1995,matsu1995,lee2000,titn2005,kd2006,blc2007}.  Recently, Ozyuzer et al. succeeded in detecting strong and continuous emission of terahertz EM waves from mesa-shaped samples of the high temperature superconductor Bi$_{2}$Sr$_{2}$CaCu$_{2}$O$_{8}$ (BSCCO) \cite{oz2007}.  The general mechanism for the emission is as follows.  When an external current is applied along the c-axis, the ac Josephson current in the resistive state excites a cavity resonance mode of Josephson plasma wave in the sample.  The excited standing wave of  Josephson plasma is converted to a terahertz EM wave at the mesa surfaces and the EM wave is emitted into the vacuum space.  However, details of the mechanism have not yet been clarified, although it is important for designing the terahertz EM wave emitters with use of IJJ.\\
\begin{figure}
\includegraphics[width = 8cm, clip]{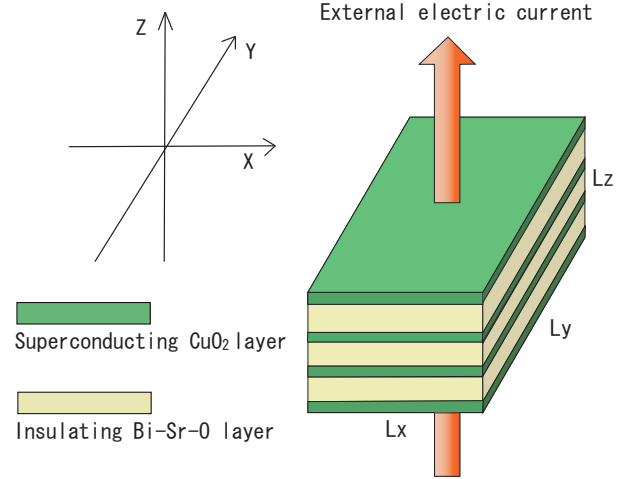}
\caption{\label{BSCCO} (color online) Schematic view of intrinsic Josephson junctions.}
\end{figure}
\indent Recently, X. Hu and S. Lin \cite{lh2008,hl2008}, and A. Koshelev \cite{kcond2008} proposed the following new mechanism.  When the inductive interaction between the superconducting CuO$_{2}$ layers in BSCCO is strong, static kink structures arise in the phase difference of superconducting order parameter between the superconducting layers.  The phase kinks induce cavity resonance modes of the Josephson plasma.  This is a new dynamic state caused by the non-linear effect special in the IJJ system.  In this paper, we first discuss the stability of this new state, and then we investigate the mechanism of the terahertz EM wave emission on the basis of the discussion.\\
\indent For the sample of IJJ, we use a model shown in Fig. \ref{BSCCO}.  In this figure the superconducting CuO$_{2}$ layers and the insulating layers in the IJJ are shown in green and light green, respectively.  An external electric current is applied in the direction of the z-axis, perpendicular to the layers.  The $L_{x}$, $L_{y}$, and $L_{z}$ are the sample lengths respectively along the $x$, $y$, and $z$-axes. Now, we derive the equation for the simulation.  The superconducting order parameter of the $l$th layer is expressed as $\Psi_{l}=\Delta_{l}(\bm{r},t)\exp[i\varphi_{l}(\bm{r},t)]$ with $\bm{r}=(x,y)$, $x$, $y$ and $t$ referring to the spatial and temporal coordinates, respectively.  We assume that the amplitude $\Delta_{l}(\bm{r},t)$ is constant independent of space and time, and only the phase $\varphi_{l}(\bm{r},t)$ is dependent on space and time.  In this case, the current density along the $z$-axis is given by a sum of the Josephson, quasiparticle, and displacement current densities as
\begin{eqnarray}
I_{l+1,l} = J_{c}\sin \psi_{l+1,l} +\sigma_{c}E_{z,l+1,l} +\frac{\epsilon}{4\pi}\partial_{t}E_{z,l+1,l}
\label{current}
\end{eqnarray}
where $J_{c}$ is the critical current density, $\sigma_{c}$ is the normal conductivity along the c axis, and $E_{z,l+1,l}$ is the electric field between $(l+1)$th and $l$th layer along the $z$ axis.  In Eq. (1), $\psi_{l+1,l}(\bm{r},t)$ is the gauge invariant phase difference defined as
\begin{eqnarray}
\psi_{l+1,l}(\bm{r},t) =&& \varphi_{l+1}(\bm{r},t) -\varphi_{l}(\bm{r},t) \nonumber\\
&&-\frac{2\pi}{\phi_{0}}\int^{z_{l+1}}_{z_{l}}A_{z}(\bm{r},z,t)dz
\label{giphase}
\end{eqnarray}
with the vector potential $A_{z}(\bm{r},z,t)$ and the flux unit $\phi_{0}$.  For the superconducting current density in the CuO$_{2}$ plane, we use the generalized London equations, since the Ginzburg-Landau parameter is very large in BSCCO.  We insert Eqs. (\ref{current}) and (\ref{giphase}) into Maxwell's equations along with the superconducting current in the CuO$_{2}$ layers.  Following the calculation procedure given in \cite{titn2005}, we have
\begin{eqnarray}
&&\partial^{2}_{x'}\psi_{l+1,l}+\partial^{2}_{y'}\psi_{l+1,l} \nonumber\\
=&& (1-\zeta \Delta^{(2)})(\sin \psi_{l+1,l} +\beta \partial_{t'}\psi_{l+1,l}+\partial^{2}_{t'}\psi_{l+1,l})
\label{Jeq}
\end{eqnarray}
In above equation, we use normalized units for length and time, respectively, defined as $x'=x/\lambda_{c}$, $y'=y/\lambda_{c}$, $t'=\omega_{p}t$, where $\omega_{p}$ is the plasma angular frequency, and $\lambda_{c}$ is the penetration depth of the magnetic field applied along the $x$ or $y$-axis from the $xz$ or $yz$-surface. The parameters in Eq. (\ref{Jeq}) are defined as $\zeta = \lambda^{2}_{ab}/sd$ and $\beta=4\pi\lambda_{c}\sigma_{c}/\sqrt{\epsilon_{c}}$, where the $s$ and $d$ are the thicknesses of the superconducting and insulating layers, respectively. $\lambda_{ab}$ is the magnetic field penetration depth from the $xy$-plane surface and $\zeta$ is the inductive constant between the CuO$_{2}$ layers. The operator $\Delta^{(2)}$ is defined as $\Delta^{(2)}f_{l}=f_{l+1}-2f_{l}+f_{l-1}$.  In Eq. (\ref{Jeq}), we neglect the charging effect in the CuO$_{2}$ layers, since we consider the region above the retrapping point, and in this region the charging effect is much weaker than the inductive effect between the CuO$_{2}$ layers.  Keeping in mind that the IJJ is Bi$_{2}$Sr$_{2}$CaCu$_{2}$O$_{8}$, we choose $\lambda_{ab}=0.4\mu$m, $\lambda_{c}=50\sim 150\mu$m, $s=3 \AA$, $d=12 \AA$, $\beta=0.02$, and $\zeta=5\times 10^{5}$.\\
\indent We express the phase difference $\psi_{l+1,l}(\bm{r},t)$ as
\begin{eqnarray}
&&\psi_{l+1,l}(\bm{r}',t') \nonumber\\
=&& \omega_{J}t' + \psi^{t}_{l+1,l}(\bm{r}',t')+\psi^{s}_{l+1,l}(\bm{r}')+\frac{I'}{4}\bm{r}'\cdot\bm{r}'
\label{phase-expansion}
\end{eqnarray}
The first term in the right hand side of Eq. (\ref{phase-expansion}) is the phase difference due to the ac Josephson effect, the second term is the phase difference due to the excited cavity mode, the third term is a static phase difference, and the fourth term is the phase difference due to an external current density normalized by $J_{c}$, $I'$.\\
\begin{figure}
\includegraphics[width = 8cm, clip]{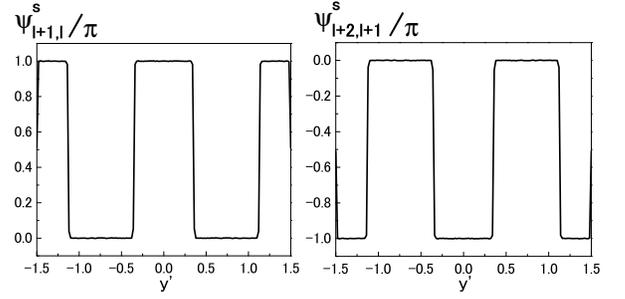}
\caption{\label{phasedif} (color online) Typical configuration of the static phase difference.}
\end{figure}
\begin{figure}
\includegraphics[width = 8cm, clip]{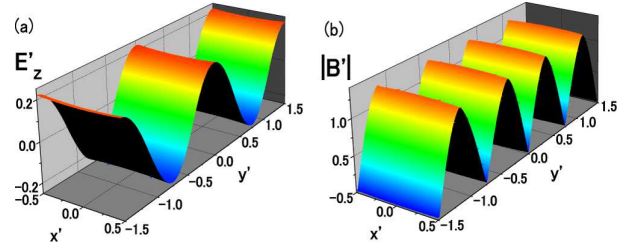}
\caption{\label{EB3D} (color online) Snapshot of the electric and magnetic fields at $V'=3.8$ and $I'=0.09$. $E'_{z}$ and $|B'|$ are electric and magnetic fields normalized by $\phi_{0}/2\pi \lambda_{c}d$. The sample with the lengths $L'x=1$ and $L'y=3$ is used.}
\end{figure}
\indent The normalized sample lengths $L'_{x}$, $L'_{y}$ and $L'_{z}$ are taken to be $1$, $3$, and $0.01$.  Although the samples used in the experiments \cite{oz2007} is composed of several hundreds of intrinsic Josephson junctions, to simplify the calculation we impose an assumption that the phase difference has a four junctions periodicity along the $z$-axis.  Since the sample length along the $z$-axis is much shorter than the wave length of terahertz EM wave, the EM waves emitted from the side surfaces of the sample are not plane waves. The boundary condition in this case has been given by N. Bulaevskii and A. Koshelev \cite{bula2006}. We use the boundary condition and tentatively take $B_{x}/E_{z} = \gamma = \pm 0.1$ at the $xz$-surfaces and $B_{y}/E_{z} = \gamma = \pm 0.1$ at the $yz$-surfaces, $E_{z}$, $B_{x}$ and $B_{y}$ being respectively the oscillating parts of electric and magnetic fields at the sample surfaces.  Under the boundary condition, we numerically solved Eq. (\ref{Jeq}) and obtained a solution of the static phase difference $\psi^{s}_{l+1,l}(\bm{r})$ with a kink and anti-kink structure at a cavity resonance voltage $V'=3.8$ as shown in Fig. \ref{phasedif}. The normalized voltage $V'$ is defined by $V/V_{p}$, $V$ and $V_{p}$ being the voltages between the CuO$_{2}$ layers and  $\hbar \omega_{p}/2e$, respectively. As seen in the figure, the kink phase structure occurs along the $y$-axis.  The oscillating electric field of the cavity mode at a time is shown in Fig. \ref{EB3D}(a). The electric field $E_{z}$ is almost uniform along the $x$-axis, and the electric field along the $y$-axis is a standing wave with two wave lengths.  The amplitude of the oscillating electric field is symmetric with respect to the center of the sample along the $x$- and $y$-axes.  The absolute value of the oscillating magnetic field $|\bm{B}|=\sqrt{B^{2}_{x}+B^{2}_{y}}$ at the time is shown in Fig. \ref{EB3D}(b).  The absolute value of $B_{x}$ is much larger than that of $B_{y}$ inside the sample and at the $yz$-surfaces.  This state with phase kinks is similar to those obtained by X. Hu and S Lin \cite{lh2008,hl2008}, and A. Koshelev \cite{kcond2008}. In the above calculation, when we took several kinds of the initial condition with modulations along the $z$-axis for the phase difference $\psi_{l+1,l}(\bm{r},t)$, we obtained the several kinds of phase kink state.  The internal energies of the phase kink states consist of superconducting current energies and electric and magnetic energies.  The internal energies have almost the same value in independence of the initial conditions.  If we took an initial condition that the phase difference is perfectly uniform along the $z$-axis, we obtained a solution of the state without phase kink in place of the solutions with phase kinks. The calculated internal energies of the phase kink states are lower than that of the state without kink phase, and the ratio of the energy of the phase kink state to the energy of the state without phase kink is about $0.8$.  Therefore, the state with phase kinks can be excited by a lower external current.  According to the facts mentioned above, it is expected that the states with phase kinks are often excited by applying external currents.\\
\indent In the calculation so far, we have assumed that the amplitude of the superconducting order parameter is spatially uniform in the CuO$_{2}$ layers.  However, the phase kinks induce a modulation of the amplitude of the order parameter as discussed in the following.  The large change of the static phase difference occurs around $y'=\pm 0.375$ and $\pm 1.125$ as seen in Fig. \ref{phasedif}.  The change causes a spatial change of the phase of the superconducting order parameter in the superconducting CuO$_{2}$ layers.  This change of the phase induces a reduction of the ampliude of the order parameter around the phase kinks.  This reduction of the order parameter amplitude is similar to that in the vortex core of the Abrikosov vortex.  According to this reduction of the amplitude, the superconducting condensation energy is reduced and thus the energy of the system is increased.  We estimate the energy increase in the following way.  Using the Ginzburg-Landau equation we calculate the order parameter $\Delta (\bm{r})$ modified by the phase kinks.  Then, the energy increase $E_{c}$ is calculated by the formula $E_{c}=(H_{c}^{2}/8\pi)\int [1-(\Delta(\bm{r})/\Delta_{0})^{2}]dV$, $H_{c}$ being the thermodynamic critical field and $\Delta_{0}$ being the amplitude of the order parameter without phase kink.  Using the experimental value of $H_{c}$ we estimate $E_{c}$ to be $1 \sim 2$ times as large as the internal energy of the state without phase kink.  If $E_{c}$ is added to the internal energy of the kink phase, the total energy of the kink state is $2 \sim 3$ times larger than the internal energy of the state without kink.  Therefore, to excite the state with phase kink, we need an external current larger than the current to excite the state without kink.  If heating is considered, the state with kink is hardly excited \cite{oz2007,yurg1997}.\\
\indent In the calculation of the resonance mode shown in Fig. \ref{EB3D}, we have missed an important fact.  As shown in Fig. \ref{EB3D} the magnitude of $B_{x}$ is much larger than that of $B_{y}$ at the $yz$- surfaces.  Although the magnetic field $B_{x}$ does not contributes to the radiation, it penetrates the vacuum from the surface.  We should add this magnetic energy in the vacuum to the internal energy of the state.  If the resonance occurs along the $x$-axis instead of along the $y$-axis, the magnitude of $B_{y}$ becomes much larger than that of $B_{x}$.  Since the $xz$-surface area is narrower than the $yz$-surface area for the sample, the magnetic energy penetrated from the $xz$-surface into the vacuum is smaller and, thus stabilizes the magnetic field $\bm{B}$ along the $y$-axis. Therefore, the cavity resonance mode actually occurs along the $x$-axis rather that along the $y$-axis in the sample \cite{oz2007}.  The origin of the magnetic energy stabilizing $\bm{B}$ along the $y$-axis in this case is similar to that of the shape anisotopy energy of a ferromagnet that stabilizes its magnetization along the longer size direction of the ferromagnet.\\
\begin{figure}
\includegraphics[width = 8cm, clip]{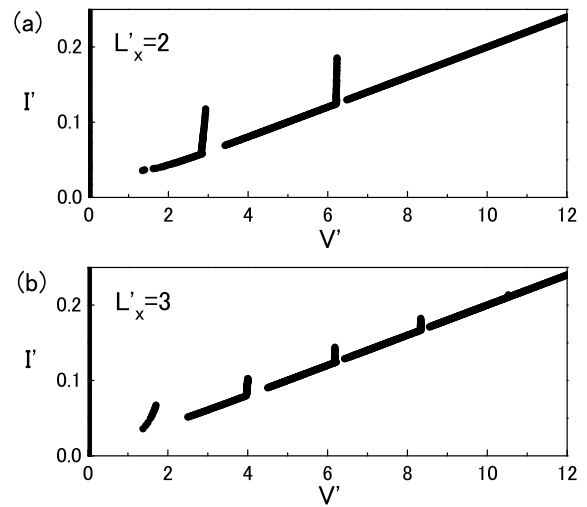}
\caption{\label{IV} (color online) I-V curves in the two cases of (a) $L'_{x}=2$ and (b) $L'_{x}=3$. The boundary condition parameter $\gamma = 0.1$ is used.}
\end{figure}
\begin{figure}
\includegraphics[width = 8cm, clip]{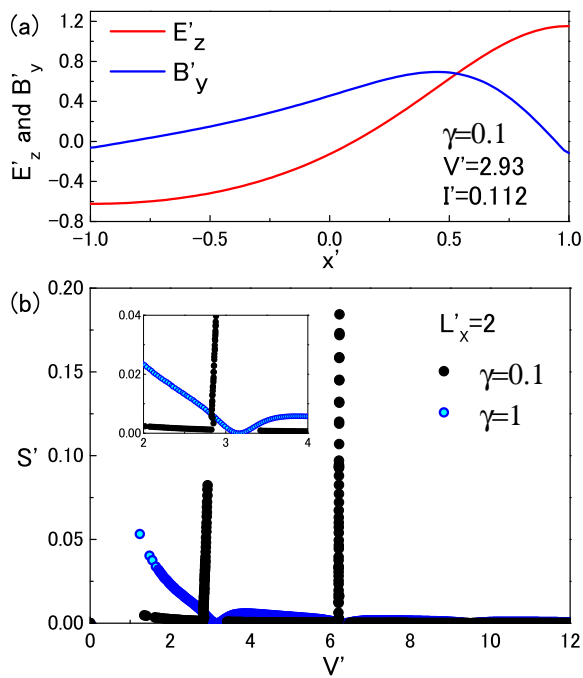}
\caption{\label{intensity} (color online) (a) Snapshot of the electric and magnetic fields at the top of first internal shapiro step $V'=2.93$ and $I'=0.112$. The normalization of $E'_{z}$ and $B'_{y}$ is the same as that in Fig. \ref{EB3D}. (b) Voltage dependence of the intensity of emission from intrinsic Josephson junctions. The length of the sample is $L'_{x}=2$. Black circles and blue circles show the emission intensities in the cases of $\gamma = 0.1$ and $\gamma =1$, respectively. The time averages of Poynting vectors are normalized by $c(\phi_{0}/2\pi \lambda_{c}d)^{2}/4\pi$, $S'$.}
\end{figure}
\begin{figure}
\includegraphics[width = 8cm, clip]{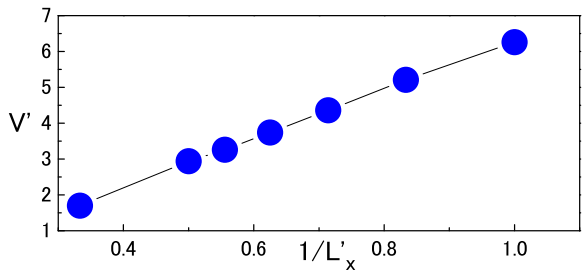}
\caption{\label{1/w} (color online) $1/L'_{x}$ dependence of the first resonance voltage.}
\end{figure}
On the basis of these discussions, we investigate the physical properties originating from the cavity resonance along the $x$-axis in the states without phase kink.  Figure \ref{IV}(a) shows the calculated current-voltage (I-V) curve for $L'_{x}=2$.  As seen in the figure, two sharp peaks appear respectively at the  normalized voltages $V'=2.9$ and $6.2$.  These sharp peaks are caused by the cavity resonance of the excited electric and magnetic fields in the sample.  Since the amplitude of the oscillating electric field is large at the resonance voltages, the sharp peaks are considered to be internal Shapiro steps induced by the oscillating electric fields in a self-consistent way \cite{wang2001}. Figure. \ref{IV}(b) shows the current-voltage curve for $L'_{x}=3$. The cavity resonance voltage at $V'=1.7$ almost coinsides with the retrapping voltage \cite{oz2007}. The oscillating electric and magnetic fields for $V'=2.9$ at a time is shown in Fig. \ref{intensity}(a).  The amplitudes of both the electric and magnetic fields are asymmetric with respect to the center of the sample.  Figure \ref{intensity}(b) shows the emission intensities (the time averages of the Poynting vectors) for the boundary condition parameters $\gamma=0.1$ and $\gamma=1$ as functions of $V'$.  The black circles indicate the emission intensities for $\gamma=0.1$.  The sharp emission peaks appear at the cavity resonance voltages as seen in Fig. \ref{IV}(a).  The maximum emission power at $V'=2.9$ is estimated to be $23$mW for the sample with $\lambda_{c}=100\mu$m. The blue circles in Fig. \ref{intensity}(b) show the emission intensities for the boundary condition parameter $\gamma=1$.  The $\gamma=1$ means that $L'_{y}$ and $L'_{z}$ are much longer than the emitted EM wave length and thus the emitted EM waves are plane waves.  As seen in the figure, the voltage dependence of the emission intensity is very different from that for $\gamma=0.1$ and the intensity vanishes at the cavity resonance voltages.  Matsumoto et al. have calculated the intensity in this case and shown that the oscillating electric and magnetic fields vanish at the $yz$-surfaces \cite{mkm2008}.  The results indicate that the emission intensity strongly depends on the sample size relative to the EM wave length. Each of the blue circles in Fig. \ref{1/w} denotes the lowest normalized voltage of the cavity resonance for each of 7 samples with different $L'_{x}$. As seen in the figure, the normalized voltage $V'$ or the normalized cavity resonance frequency $\omega'$ is inversely proportional to $L'_{x}$ in agreement with the experiments \cite{oz2007}.\\
\indent The authors thank X. Hu, A. Koshelev, L. Bulaevskii, M. Matsumoto, M. Machida, S. Lin, K. Kadowaki, 
U. Welp, K.E.Gray, L. Ozyuzer, W.K.Kwok, C. Kurter,and H. B. Wang for valuable discussions.  One. of the authors (M.T.) thanks H. Koinuma, H. Takagi, K. Itoh, and K. Itaka for great support to this research.  Authors (M.T. and S. F) thank Y. Takada for giving us the opportunity of collaboration. This work has been supported by the JST (Japan Science and Technology Agency) CREST project and by the CTC (Core-to-Core) program.


\end{document}